# Thermoelectric Transport Coefficients in Mono-layer MoS$_2$ and WSe$_2$: Role of Substrate, Interface Phonons, Plasmon, and Dynamic Screening

Krishnendu Ghosh†, and Uttam Singisetti*

*Electrical Engineering Department, University at Buffalo, Buffalo, NY, 14260, USA*

**Abstract:** The thermoelectric transport coefficients of electrons in two recently emerged transition metal di-chalcogenides (TMD), MoS$_2$ and WSe$_2$, are calculated by solving Boltzmann Transport equation and coupled electrical and thermal current equations using Rode's iterative technique. Scattering from localized donor impurities, acoustic deformation potential, longitudinal optical (LO) phonons, and substrate induced remote phonon modes are taken into account. Hybridization of TMD plasmon with remote phonon modes is investigated. Dynamic screening under linear polarization response is explored in TMDs sitting on a dielectric environment and the screened electron-phonon coupling matrix elements are calculated. The effect of screening and substrate induced remote phonon mediated scattering on the transport coefficients of the mentioned materials is explained. The transport coefficients are obtained for a varying range of temperature and doping density for three different types of substrates – SiO$_2$, Al$_2$O$_3$, and HfO$_2$. The thermoelectric properties of interest including Seebeck coefficient, Peltier coefficient, and electronic thermal conductivity are calculated.

†Email: kghosh3@buffalo.edu, Tel: 716-645-1017

*Email: uttamsin@buffalo.edu, Tel: 716-645-1536, Fax: 716-645-3656

## I. INTRODUCTION

Transition metal di-chalcogenides (TMD) have recently attracted unprecedented attention [1-10] in the electron device community especially as a replacement of silicon for transistor channels. Unlike graphene, these materials have an intrinsic band-gap which makes them suitable for transistor channels reminiscent of an ideal switch for digital logic applications. Among the TMDs, $MoS_2$ has been investigated extensively in particular due to its abundance in nature [1-6]. While, $WSe_2$ has got attention [7-10] due to its intrinsic doping property which allows the scope for observance of ambipolar characteristic [9]. However, in addition to be studied as potential transistor channel semiconductor, these materials have interesting mechanical, optical, and thermoelectric properties that could be used for various emerging applications [6]. Thermoelectric properties of these materials in particular have significant technological importance due to their capability of converting thermal energy to electrical energy and vice-versa [11-15]. Important thermoelectric properties of interest include Seebeck coefficient ($\alpha$), Peltier coefficient ($\pi$), and thermal conductivity ($\kappa_e$). Though extensive research has been carried out on the electronic properties of TMDs, their thermoelectric properties have not been studied much. Recent *ab initio* calculations [16-18] have shown the thermoelectric properties of the TMDs which considered either ballistic transport or a constant energy independent mean free path. However, these assumptions are not consistent with the measured transport properties of the TMDs. Indeed a strong evidence of diffusive transport is reflected in the low carrier mobility of these materials. Moreover, Rode [19] has shown that the Seebeck coefficient does not depend on the absolute value of the scattering rates, rather it depends on how the scattering rate depends on electron energy. Detailed quantitative calculations and careful analysis here has shown that it is important to consider the effect of scattering as they do impact the thermoelectric properties significantly. A paramount role is played by the substrate on which the TMD is grown or transferred. Like graphene [20], TMDs also suffer from scattering mediated by remote optical phonon modes of the substrate. On one side, high-k substrates suppress the Coulomb scattering [21, 22], while on the contrary, their strong TO modes cause additional scattering [22]. Similar to mobility calculations, this trade-off becomes important for thermoelectric transport coefficients too.

In this paper, we solve the Boltzmann transport equation (BTE) and coupled electrical and thermal current equations for $MoS_2$ and $WSe_2$ to extract thermoelectric transport coefficients. Scatterings from donor impurities, longitudinal optical (LO) polar modes, acoustic deformation potential, and substrate induced remote phonon modes are taken into account. We use Rode's iterative method in order to take into account the inelastic and anisotropic nature of the electron-LO mode scattering and also electron-remote phonon mode scattering [23, 24]. A dynamic screening model is used assuming random phase approximation. The organization of the paper is as follows. In Section II we start with the procedure used to solve the BTE and coupled current equations. In Section III we briefly discuss the models for elastic scattering mechanisms and the Frölich electron-LO mode interaction. In Section IV we discuss the formation of interface plasmon-phonon (IPP) modes due to coupling of plasmon with the remote phonon modes. Here, we also talk about calculation of corresponding scattering field strength and matrix elements. Section V covers the details of screening calculation including discussion on Landau damping. In Section VI we give insight on our results and findings of this study followed by conclusion in Section VII.

## II. THE BTE AND COUPLED CURRENT EQUATIONS

The electrical current density due to flow of electrons is given by the contribution from the non-symmetric part of the electronic distribution function:

$$J_x = -\frac{e}{\Omega}\Sigma_k v_x(\boldsymbol{k})f_A(\boldsymbol{k}) \tag{1}$$

Here, $v_x$ is the group velocity of electrons in the transport direction, $f_A(\boldsymbol{k})$ is the non-symmetric part of the electronic distribution which is discussed later, $\boldsymbol{k}$ is the electron wave-vector, and $\Omega$ is a normalization area. On the other hand, the heat current density due to electron flow is given by the kinetic energy flow contribution from the non-symmetric part of the electronic distribution function:

$$J_{Qx} = -\frac{e}{\Omega}\Sigma_k [E(\boldsymbol{k}) - E_F] v_x(\boldsymbol{k})f_A(\boldsymbol{k}) \tag{2}$$

Here, $E$ is the electron energy and $E_F$ is the chemical potential. We calculate the non-symmetric part of the electronic distribution function using Rode's iterative calculation [19]. Under low electric fields the entire distribution function can be written as

$$f(\boldsymbol{k}) = f_0(k) + g(k)\cos\theta \tag{3}$$

Here, $f_0$ is the equilibrium Fermi-Dirac distribution and $\theta$ is the angle between the applied electric field direction and $\boldsymbol{k}$. The point to be noted here is that $g$ only depends on the magnitude of the electron wave-vectors. Rode's iteration solves for $g$ using the following conformal mapping [19],

$$g_{i+1} = \frac{S_{in}(g_i') - \frac{eF\partial f_0(k)}{\hbar \partial k} - v\frac{\partial f_0}{\partial x}}{S_{out}(k) + 1/\tau_m(k)} \tag{4}$$

Here, $S_{out}$ and $S_{in}$ are the net out-scattering rate and in-scattering rate respectively. $\tau_m$ is the momentum relaxation time for the elastic scattering processes discussed in Section III. $F$ is the applied parallel electric field, which should be low enough for Eq. (3) to be good. $\frac{\partial f_0}{\partial x}$ can be expressed in terms of $\frac{\partial T}{\partial x}$ (which is done in [19] for a bulk system and in Eq. C8 of this work for a 2D system) For calculating $S_{out}$ and $S_{in}$ from the scattering matrix elements we follow the same procedure described in our previous work for a 2DEG system [24]. The iteration (Eq. (4)) begins with a $g_0$ given by the relaxation time approximation. Once the distribution converges, $J_x$ and $J_{Qx}$ can readily be calculated from Eq. (1) and Eq. (2) respectively.

Fig. 1 shows the outline of the entire calculation. First we consider the device under isothermal condition. We calculate the drift mobility from scattering matrix elements and store it for later use. Next we apply a small temperature gradient ($\frac{\partial T}{\partial x} = 10^3$ K/m) with the device ends being electrically shorted and compute the corresponding spatial gradient of the Fermi-Dirac distribution function. Using this and the previously calculated scattering matrix elements we run Rode's iteration again. Please note here, since the two ends are shorted the second term on the numerator of Eq. (4) vanishes ($F = 0$). Once the Rode's iteration converges the electric current density ($J_x$) and the heat current density ($J_{Qx}$) can be calculated from which the transport coefficients can be extracted from their definition [25]

$$J_x = \sigma_x \frac{\partial E_F}{\partial x} - \frac{B_x}{T^2}\frac{\partial T}{\partial x} \tag{5a}$$

$$J_{Qx} = p_x \frac{\partial E_F}{\partial x} - \frac{K_x}{T^2}\frac{\partial T}{\partial x} \tag{5b}$$

Here, $\sigma_x, B_x, p_x$, and $K_x$ are the transport coefficients. An important point to be noted here is that we assume an isotropic crystal in two dimensions so that the transport coefficient tensors are actually diagonal. The isothermal calculations store the values of $\sigma_x$ and $p_x$. The remaining two coefficients can be extracted from the $J_x$ and $J_{Qx}$ calculated with the temperature gradient.

The thermoelectric properties of interest that can be computed from here are the Seebeck coefficient, Peltier coefficient, and thermal conductivity (electronic contribution). They are related to the transport coefficients by the following relations [25]:

$$\alpha = \frac{B_x}{\sigma_x T^2}, \qquad \pi = \frac{p_x}{\sigma_x}, \qquad \kappa_e = \frac{K_x - \frac{p_x B_x}{\sigma_x}}{T^2}$$

$\alpha$, $\pi$, and $\kappa_e$ are the Seebeck coefficient, Peltier coefficient, and thermal conductivity respectively.

## III. INTRINSIC SCATTERING MECHANISMS

Several intrinsic scattering mechanisms as discussed in the introduction make the transport diffusive in TMDs. We consider impurity and acoustic deformation potential scattering matrix elements as were considered in references [22] and hence we do not describe them here. For the impurity Coulombic scattering we consider ionized donor scattering ($N_D = n_S$, where $n_S$ is the electron concentration and $N_D$ is the ionized donor impurity concentration) without any additional impurities. The momentum relaxation times from each of these two types of scattering mechanism are calculated and are plugged in the Rode's iteration loop (Eq. 4) as the elastic scattering parameter. Another very important scattering mechanism originates from the electron – LO phonon interaction. However, the conventional macroscopic models do not hold good for atomistically thin semiconductors as pointed out by Kaasbjerg *et al*. [26]. We first follow their analytic approach (Eq. 9 of [26]) to find the polar LO phonon scattering rate. Next, we employ dynamic screening on the electron-LO phonon coupling matrix to improve the accuracy of the calculation; the dynamic screening behavior is discussed in Section V.

## IV. SCATTERING BY REMOTE INTERFACE PHONONS

The presence of remote phonon modes at the interface of a dielectric and a semiconductor has been studied extensively for 3D and 2D semiconductors [27-31]. But coupling of such remote phonon modes with the plasmon was not discussed for TMDs, though it is discussed in details for graphene and 2DEGs [20, 32] . We consider a simple geometry, relevant for thermoelectric applications, shown in the inset of Fig. 2(a) where a TMD material is sitting on a substrate and the other side of the interface is air.

### A. COUPLED PLASMON-REMOTE PHONON MODES

In such a system the potential due to the remote phonon modes can be written as,

$$\varphi_q^\nu(z) = A_{q,\omega_q^\nu} exp(-qz) ; z > 0$$

$$= B_{q,\omega_q^\nu} exp(qz) ; z < 0 \quad (6)$$

$q, \omega_q^\nu$ represent the wave-vector and energy of the phonon field with $\nu$ being the mode index. Implementing Dirichlet and Neumann boundaries we obtain the secular equation,

$$\varepsilon_{TMD}(q, \omega_q^\nu) + \varepsilon_{sub}(q, \omega_q^\nu) = 0 \quad (7)$$

Here, $\varepsilon_{TMD}(q, \omega_q^\nu)$ and $\varepsilon_{sub}(q, \omega_q^\nu)$ are the two out-of plane dynamic dielectric constants of the TMD and the substrate respectively. In long-wavelength limit they have the following form:

$$\varepsilon_{TMD}(q, \omega_q^\nu) = \varepsilon_{TMD}^\infty \left(1 - \frac{\omega_P^2}{\omega^2}\right) \quad (8a)$$

$$\varepsilon_{sub}(q, \omega_q^\nu) = \varepsilon_{sub}^\infty + \frac{\varepsilon_{sub}^0 - \varepsilon_{sub}^{int}}{\omega_{TO,1}^2 - (\omega_q^\nu)^2} \omega_{TO,1}^2 + \frac{\varepsilon_{sub}^{int} - \varepsilon_{sub}^\infty}{\omega_{TO,2}^2 - (\omega_q^\nu)^2} \omega_{TO,2}^2 \quad (8b)$$

$\varepsilon_i^\infty$, and $\varepsilon_i^0$ are the high frequency and static dielectric constants of the corresponding materials. $\varepsilon_{sub}^{int}$ is an intermediate dielectric constant defined to distinguish the contributions of the two TO phonon modes to the net dielectric constant of the substrate. $\omega_P$ is the plasmon energy of the TMD which is discussed in [33]. $\varepsilon_{TMD}^\infty$ is the high frequency dielectric constant of the TMD. Detailed discussion on the TMD dielectric constant and plasmon vibration is done later in Section V. $\omega_{TO,1}$ and $\omega_{TO,2}$ are the TO mode energies in the substrates. The list of substrate parameters is given in Table I, while Table II lists the materials parameters used for MoS$_2$ and WSe$_2$. Solving the secular equation we obtain the dispersion relation for the interface plasmon-phonon (IPP) modes for MoS$_2$ on HfO$_2$ as shown in the main panel of Fig. 2(a). It can be seen that there is significant coupling between the surface modes and the plasmon mode for the bottom two modes which will impact the electron scattering rate.

B.  PHONON CONTENT AND SCATTERING STRENGTH

Since only the phonon part of the IPP coupled modes effectively causes exchange of energy between surface modes and electrons [34], we need to separate out the phonon contributions from these coupled modes. The phonon content is extracted following the method described in [20, 24, 32, 34]. The phonon content of the three modes in Fig. 2 (a) is plotted in Fig. 2(b) ($\Phi^{\nu,TOi}$, $TOi$ is the $i^{\text{th}}$ TO mode of the substrate).

Having obtained the dispersion relation, and the phonon content of the coupled IPP modes we calculate [30] the coefficient $A_{q,\omega_q^\nu}$ in Eq. (6) ($= B_{q,\omega_q^\nu}$, from Dirichlet boundary condition at $z = 0$) by equating the electrostatic energy of the scattering field with the ground state quantum mechanical energy of the modes

$$\int \varphi_q^\nu(z)\rho_q^\nu(z)dz = \frac{1}{2}\hbar\omega_q^\nu \tag{9}$$

Here $\rho_q^\nu$ is the polarization charge induced by the scattering field associated with the phonon at $(q, \omega_q^\nu)$. The polarization charge can be obtained by solving the Poisson's equation with a tailored dielectric function [30]. By tailored dielectric function we mean that we want to separate out the contribution of each phonon mode in inducing that polarization charge. This can be done by separating out each phonon contribution in the dielectric function of the substrate. The details of this calculation is given in Appendix A. Here we give the final form of the phonon contributed screened scattering field strength for each mode,

$$\Lambda^{\nu,TOi}(q)$$

$$= \sqrt{\frac{\hbar\omega_q^\nu}{\varepsilon_0 q}\left(\frac{1}{\varepsilon_{sub}^{\nu,-TOi}(q,\omega_q^\nu) + 1 + \frac{e^2\Pi(q,\omega)}{2q}\phi_1} - \frac{1}{\varepsilon_{sub}^{\nu,+TOi}(q,\omega_q^\nu) + 1 + \frac{e^2\Pi(q,\omega)}{2q}\phi_1}\right)}\Phi^{\nu,TOi}$$

$$\tag{10}$$

which looks analogous to but simpler than what was derived for top gated graphene device in [32]. Here, $\varepsilon_{sub}^{v,-TOi}$ and $\varepsilon_{sub}^{v,+TOi}$ are the mentioned tailored dielectric functions discussed in Appendix A. The term $\frac{e^2\Pi(q,\omega)}{2q}\phi_1$ comes from electronic polarization inside the TMD which is discussed in section V. The scattering strengths calculated this way is plotted (normalized by $\sqrt{\frac{\hbar\omega_q^v}{\varepsilon_0 q}}$) in Fig. 2(c).

### C. ELECTRON-IPP COUPLING MATRIX AND SCATTERING RATE

Once the scattering field strength is obtained, calculation of the coupling matrix elements with the electrons is straightforward. The interaction Hamiltonian can be written as,

$$H_q^v = \sqrt{\sum_{i=1,2}\left(\Lambda^{v,TOi}(q)\right)^2}\left(e^{iq\cdot r}a_q^{v+} + e^{-iq\cdot r}a_q^v\right) \quad (11)$$

Here, $a_q^{v+}$ and $a_q^v$ are the IPP creation and annihilation operators. To calculate the electron-IPP coupling matrix, for simplicity, we take the out-of-plane electronic wave function to be sinusoidal and compute the matrix elements as

$$M_q^v = \frac{2}{t}\int_0^t H_q^v \exp(-qz)\sin^2\left(\frac{\pi z}{t}\right)dz \quad (12)$$

$t$ is the thickness of the TMD, taken to be 5Å in this work. The retardation effects [27] at very small wave-vectors ($q \approx \frac{\omega_{TO}}{c}$, $c$ is the speed of light in that medium) are safely ignored. After the matrix elements are calculated they can be plugged into the Fermi Golden rule to get the scattering probabilities ($P(k, k')$) from an electronic state $k$ to $k'$. These scattering probabilities are used in the out-scattering equation of our previous work (Eq. 19(b) of [24]). The calculated out-scattering rate $S_{IPP}$ is plotted in Fig. 2(d) for two different electron densities ($n_s$) and also for comparison the calculated impurity mediated momentum relaxation rate ($S_{IMP}$) is plotted. The sharp kinks in $S_{IPP}$ arise due to onset of phonon emission while the tiny jitters come due to numerical gridding of the $q$ space. The essential point to note here, which will be crucial while analyzing thermoelectric properties, is that at low energy ranges the impurity scattering rate swiftly drops well below the phonon scattering rate. Also, as we increase the electron densities the impurity scattering rate increases due to more number of donor impurities (Note: $N_D = n_s$ here). The dependence of phonon scattering rate on $n_s$ is an interplay between enhanced screening, which tries to cut down the rate, and an expanding Fermi radius, which tries to increase the rate. For this particular case of $HfO_2$, enhancement of screening with carrier density is not pronounced due to a large dielectric mismatch between the substrate and the TMD. Hence the scattering rate increases slightly with $n_s$.

## V. DYNAMIC SCREENING

We include dynamic scattering with finite temperature polarizability in the scattering rate calculations following the random phase approximation [33, 35]. The dynamic dielectric function can be obtained from polarization as

$$\varepsilon_{TMD}(q,\omega) = \varepsilon_{TMD}^{\infty}(1 + \frac{e^2 \Pi(q,\omega)}{2\varepsilon_{TMD}^{\infty} q}(\phi_1 + \phi_2)) \tag{13}$$

Where ($\Pi(q,\omega)$) is the dynamic polarizability. $\phi_1$ and $\phi_2$ are form factor and dielectric mismatch factor respectively defined in [36]. The dielectric function calculated this way is shown in the contour plot shown in the inset of in Fig. 3(a). The main panel of Fig. 3(a) shows the comparison between a Thomas-Fermi like static screening and a dynamic screening which clearly shows the necessity of taking dynamic screening at long wave-length limit which is explained further in the next paragraph. In the long-wavelength limit, the TMD dielectric function given in Eq. (13) can be expressed as the one given in Eq. (8a).

Plasmons can absorb (emit) energy from (to) an external electro-magnetic field which leads to damped oscillation, known as Landau damping. Fig. 3(b) shows the region of plasmon damping due to the interaction of it with an external field (known as single particle excitation (SPE)). The emission and absorption boundaries of SPE regions are calculated from momentum and energy conservation in a 2D system. If we observe the plot in Fig 3(a) we see that the effect of plasmon has a significant role in the dynamic screening process. In the long-wavelength limit when the plasmon energy lies below the phonon energy (shown as $\omega = 0.015\ eV$ in Fig. 3(a), $\omega_p=$ ) and with increasing wave-vector effective dielectric constant drops due to *anti-screening* effect by the plasmon, however as the plasmon energy approaches the phonon energy the anti-screening effect decreases and screening tends to begin. Finally, the onset of SPE cuts down the effect of the plasmon and at large wave-vectors the normalized dielectric constant approaches unity. On the other hand, the static screening limit assumes a fully responsive plasmon outside SPE region and thus the anti-screening effect is not captured.

## VI. RESULTS AND DISCUSSIONS

Having set-up the theoretical details of our calculation, the transport coefficients and the corresponding thermoelectric figure-of-merits are calculated for a temperature range of 50K – 500K keeping the density of electrons within the TMD fixed at $10^{11}$ /cm$^2$ and similarly the same are calculated for an electron density range of $10^{11} – 10^{13}$/cm$^2$ with the temperature fixed at 300K. All these calculations are done for two different TMDs, MoS$_2$ and WSe$_2$, each with three different substrates (SiO$_2$, Al$_2$O$_3$, and HfO$_2$). While mobility is extracted from isothermal device, thermoelectric transport coefficients are calculated for a temperature gradient of 1K/mm. The computational details and convergence rate of the calculation is given in Appendix B. We discuss the results below.

### A. MOBILITY

The calculated mobility values are plotted in Fig 4(a) – 4(d). Fig. 4(a) and 4(c) show the effect of temperature for MoS$_2$ and WSe$_2$ respectively while Fig. 4(b) and 4(d) show the corresponding effects of electron density at 300 K. It can be seen from Fig. 4(a) and 4(c) that as the temperature is increased the mobility on SiO$_2$ supported TMD increases initially which is a signature of the donor impurity scattering limited mobility. Then the mobility starts dropping from around 200 K which signify two important things. First is the onset of the remote phonon scattering mediated by the TO mode of SiO$_2$ that occurs around 55 meV and the second crucial point is the degradation of the polarizability of the TMD electrons due to thermal randomization. Now if we take a look at the mobility values for TMD supported by Al$_2$O$_3$ we find that the mobility at low temperature is more than an order higher than that SiO$_2$ which is a characteristic of better dielectric environment screening of impurity scattering. The mobility starts dropping around 150K due to the onset of remote phonon scattering by the lower TO mode of Al$_2$O$_3$ (48 meV). At higher temperature we see that mobility drops more rapidly compared to that in case of SiO$_2$ which is because of the nearby second TO mode of Al$_2$O$_3$ at 71 meV while SiO$_2$ has the second TO mode at a very high energy at 138 meV. The degradation of polarizability with temperature affects the overall screening less in case of Al$_2$O$_3$ than that in SiO$_2$ because of a higher dielectric mismatch (difference between dielectric constant of substrate and TMD) in case of Al$_2$O$_3$. Finally we focus on HfO$_2$ supported TMD. Here, we do not see any sign of donor impurity scattering for strong dielectric environment screening and also for the very early onset of remote phonon scattering by the lower TO mode of HfO$_2$ at 12 meV.

Next we discuss the electron density dependence. From Fig. 4(b) and 4(d) we see that with increasing n$_s$ SiO$_2$ supported TMD shows decline in mobility due to enhanced impurity scattering (note $N_D = n_s$) while HfO$_2$ shows slight improvement in mobility due to enhanced screening by electrons. Al$_2$O$_3$ shows non-monotonic trend due to a strong interplay between remote-phonon scattering and impurity scattering.

## B. THERMOELECTRIC COEFFICIENTS

The calculated Seebeck coefficients for MoS$_2$ and WSe$_2$ under different temperature and different electron density are plotted in Fig. 5(a)-(d). The Seebeck coefficient in a non-degenerate 2D material can be expressed under RTA (see Appendix C) as $-\frac{k_B}{e}\left(2 - r - \frac{E_F}{k_B T}\right)$, where $r$ is the energy exponent in the scattering rate. The low temperature Seebeck coefficients obtained from the analytical expression agrees well to the detailed calculation using Rode's method due to the reduced phonon scatterings (both LO and IPP) at low temperature, it deviates at higher temperatures when the phonon scattering rate increases. However, an important message, similar to that discussed before for bulk materials [19], to note from the given expression is that the Seebeck coefficient depends on how the scattering rate depends on energy and not on the absolute values of the scattering rates. As can be seen from Fig. 2(d) impurity scattering at lower energies drop swiftly with energy. A simple fitting gives a $E^{-1.1}$ roll off, while at a similar energy range IPP (and also LO) phonon scattering rate depends weakly on energy. The comparison between the analytical calculation and the computed data for SiO$_2$ is shown in Fig. C.2. At higher temperature $r$ deviates from -1.1 because of the onset of phonon emission process and temperature dependent polarizability and more importantly RTA becomes invalid. To discuss the effect of substrates at low temperature most of the electrons stay near the edge of the conduction band where the dominant scattering mechanism is the one mediated by impurities. Hence, as can be seen from Fig. 5(a) and 5(c), at low temperature SiO$_2$ can produce a better Seebeck coefficient while HfO$_2$ being a source of dominant IPP scattering (even at low temperature from the lower TO mode of HfO$_2$) produces low Seebeck coefficient. The cross-over near room temperature clearly reflects the temperature dependent influence of the two opposite substrates (in terms of their dielectric constant) in bringing out an optimum thermoelectric power from a TMD.

Fig. 5(b) and (d) show the dependence of Seebeck coefficient on density. Increasing density reduces the Seebeck coefficient which is well known. Our goal in this paper is to study the effect of the substrates. To get an intuitive insight we can crudely rewrite the analytical expression for Seebeck coefficient as $-\frac{k_B}{e}\left(2 - r - \ln(\frac{\pi\hbar^2 n_S}{2m^* k_B T})\right)$. At low value of $n_S$ the logarithmic term dominates (with a negative value) and all substrates produce similar Seebeck coefficient. However as we increase $n_S$ and approach $10^{13}$ the logarithmic term tends to vanish and the Seebeck coefficient is determined by $r$ which depends on the substrate we choose. Hence $SiO_2$ shows a higher Seebeck coefficient due to impurity dominated scattering. The corresponding Peltier coefficient for each case are plotted in Fig. 6(a)-(d). The effect of substrate on Peltier coefficient is less compared to that on Seebeck coefficient since by definition Peltier coefficient is a ratio of $p_x$ and $\sigma_x$, both of which have similar effect from the substrates.

## C. THERMAL CONDUCTIVITY

Here, we calculate the electronic contribution to the thermal conductivity. As can be seen from the definition of thermal conductivity in terms of the transport coefficients in section II, a high value of $\sigma_x$ and $K_x$, while a low value of $p_x$ and $B_x$ are preferred to attain a high electronic thermal conductivity ($\kappa_e$). $p_x$ and $B_x$ can both be shown to be the first order moment of the non-equilibrium part of the distribution function [25] (in our case that is $g$, the output from Rode's iteration) about the chemical potential. $\sigma_x$ and $K_x$ are the corresponding zero order and second order moments respectively. So the magnitude of thermal conductivity strongly depends on the nature of $g$ which in turn is governed by the substrate. A strongly peaked $g$ near the bottom of the conduction band (more scattering) reduces all the four coefficients, while a relatively flatter $g$ (low scattering) increases all the coefficients and hence reduce $\kappa_e$. This is why $\kappa_e$ will have a maxima for moderate inelastic scattering rates. This phenomenon can be clearly seen in Fig. 7(a) and 7(c) where Al$_2$O$_3$ produces the highest $\kappa_e$ because of an intermediate inelastic scattering rate between SiO$_2$ and HfO$_2$. The cross-over of the curves for SiO$_2$ and HfO$_2$ reflects that the corresponding electronic distribution function follows an opposing trend (from a sharply peaked one to a flatter on for SiO$_2$ and the opposite way for HfO$_2$) as we increase the temperature.

Next we take a look at the density dependence of $\kappa_e$ shown in Fig. 7(b) and 7(d). The trend of $\kappa_e$ with carrier density is a result of an interplay between mobility and carrier density. For SiO$_2$, it shows a decline in mobility with increasing $n_s$ (Fig. 4(c-d)) but the increasing $n_s$ tries to increase the conductivity. Similarly, HfO$_2$ shows a mobility that is relatively independent of $n_s$ and hence with increase of $n_s$ HfO$_2$ shows improving thermal conductivity. Al$_2$O$_3$ once again shows a clear non-monotonous behavior due to shifting of scattering mechanism from phonons to impurities. An important point to mention here is that the absolute values of $\kappa_e$ reported in this work is for a TMD of thickness 5Å and unlike, mobility and Seebeck coefficients, $\kappa_e$ depends on thickness. However, the trends with temperature, density, and substrate are expected to be unaffected.

## D. THERMO-ELECTRIC FIGURE-OF-MERIT

The thermoelectric figure-of-merit [16] defined as $ZT = \frac{\alpha^2 \sigma T}{\kappa_e + \kappa_{ph}}$ is plotted in Fig. 8(a) and 8(b) at room temperature under varying electron density for different substrates. Here, $\sigma$ is the electrical conductivity and $\kappa_{ph}$ is the lattice contribution to thermal conductivity which we took to be recently reported value [37] of 19.5 W/K.m. Comparing the magnitude of our calculated ZT with previously reported values from ballistic transport calculation we find that our ZT values are significantly low [16]. The reason for this is the low electrical conductivity obtained in our calculations due to the consideration of a highly diffusive transport. This argument can be verified if we take a look at the trend of ZT with electron density shown in Fig. 8(a-b). HfO$_2$ shows a significant improvement in ZT with increasing $n_s$ which is a result screening of phonon modes and absence of adequate impurity scattering due to strong dielectric environment screening. While $SiO_2$ and $Al_2O_3$ do not show such improvement since increasing $n_s$ screens the phonon modes well but at the same time enhanced impurity scattering does not let $\sigma$ increase. To further verify that it is the diffusive transport that kills ZT we carry out the calculation with a very low impurity concentration ($N_D = 10^{10} /cm^2, n_s = 10^{13}/cm^2$ ) for both $SiO_2$ and $HfO_2$ at room temperature. While $SiO_2$ shows a 50 times improvement in ZT, $HfO_2$ shows negligible effect due to IPP dominated scattering. So the application of a gate voltage on $SiO_2$ to increase $n_s$ with low trap density will improve ZT.

## VII. CONCLUSION

We have reported a theoretical calculation of thermoelectric transport coefficients of two TMDs, $MoS_2$ and $WSe_2$, taking into account a diffusive transport mechanism. The importance of taking into account a diffusive transport mechanism can be realized in the huge influence of substrates on the transport coefficients. Effect of substrates is analyzed quantitatively with several aspects including plasmon-remote phonon coupling and dynamic screening. Even in absence of additional impurities the mobility of a TMD remains a bottleneck due to substrate induced remote phonon scattering in a high-dielectric environment. An important issue to discuss is how the Seebeck coefficients of the TMDs compare with conventional materials like GaAs. Reference [38] shows measured Seebeck coefficient for GaAs for an n-type doping of 7.7 x $10^{18}$ /cm$^3$ is around 100 µV/K at room temperature while a bulk SiGe alloy [39] also shows Seebeck coefficient of around 100 µV/K at room temperature while a corresponding nano composite material $Si_{0.80}Ge_{0.20}B_{0.016}$ shows ~160 µV/K. On the other hand, in this work, $MoS_2$ at room temperature showed a Seebeck coefficient of 500~550 µV/K depending on the substrate chosen for a doping density of $10^{11}$ /cm$^2$ (a bulk analogue of this is roughly 2 x $10^{18}$ /cm$^3$ for a sample thickness of 5Å)

This increase is primarily due to an increased density of states at lower electron energies as pointed out for a quantum well structure [40]. The low thermal conductivity values are also promising for thermoelectric applications, although ZT has a low value due to diffusive nature of the transport. A possible improvement of ZT in a low-dielectric environment can be made by application of a top-gate voltage. Hence, to conclude, the TMDs produce a better thermoelectric performance compared to other conventional materials, however selection of a proper substrate is crucial based on other physical conditions like temperature and carrier density to optimize performance.

**APPENDIX A: SCATTERING STRENGTH OF THE IPP MODES**

The decaying surface modes are given by Eq. (6). Applying Poisson's equation we find out the corresponding polarization charge as

$$\rho_q^\nu(z) = \varepsilon_{sub}(q,\omega_q^\nu)A_{q,\omega_q^\nu}q^2 e^{qz} + \varepsilon_{TMD}(q,\omega_q^\nu)A_{q,\omega_q^\nu}q^2 e^{-qz} \quad (A1)$$

Plugging this and Eq. (6) in Eq. (9) we get the mode amplitude

$$A_{q,\omega_q^\nu} = \sqrt{\frac{\hbar\omega_q^\nu}{q\left(\varepsilon_{sub}(q,\omega_q^\nu)+\varepsilon_{TMD}(q,\omega_q^\nu)\right)}} \quad (A2)$$

But the dispersion relation in Eq. 7 gives, $\varepsilon_{TMD}(q,\omega_q^\nu) + \varepsilon_{sub}(q,\omega_q^\nu) = 0$ which leads the mode amplitude to infinity. This problem arises since we have not separated out the contribution of the phonon modes in the dielectric constant of the substrate. There comes the necessity of defining the so-called 'tailored' dielectric functions. The function $\varepsilon_{sub}^{\nu,-TOi}(q,\omega_q^\nu)$ in Eq. 10 is the dielectric function of the substrate with the substrate phonon mode $TOi$ being frozen and is defined as [30],

$\varepsilon_{sub}^{\nu,-TOi}(q,\omega_q^\nu) = \varepsilon_{sub}^\infty \frac{\omega_{LO,j}^2-(\omega_q^\nu)^2}{\omega_{TO,j}^2-(\omega_q^\nu)^2}$ $(i \neq j)$ and the function $\varepsilon_{sub}^{\nu,+TOi}(q,\omega_q^\nu)$ gives the dielectric function of the substrate with the substrate phonon mode $TOi$ being in 'full' response and is defined as $\varepsilon_{sub}^{\nu,+TOi}(q,\omega_q^\nu) = \varepsilon_{sub}^\infty \frac{\omega_{LO,j}^2-(\omega_q^\nu)^2}{\omega_{TO,j}^2-(\omega_q^\nu)^2}\left(\frac{\omega_{LO,i}^2}{\omega_{TO,i}^2}\right)$. Hence the contribution of the phonon mode

mode $TOi$ to the net mode amplitude is given as the difference between these two cases [30]

$$A_{q,\omega_q^\nu} = \sqrt{\frac{\hbar\omega_q^\nu}{\varepsilon_0 q}\left(\frac{1}{\varepsilon_{sub}^{\nu,-TOi}(q,\omega_q^\nu)+\varepsilon_{TMD}(q,\omega_q^\nu)} - \frac{1}{\varepsilon_{sub}^{\nu,+TOi}(q,\omega_q^\nu)+\varepsilon_{TMD}(q,\omega_q^\nu)}\right)} \quad (A3)$$

Multiplying the square of the mode amplitude with phonon content of each mode gives the square of the scattering strength which is Eq. 10 in the main text.

## APPENDIX B: CONVERGENCE OF RODE'S ITERATIVE METHOD FOR 2D SCATTERING RATES

Eq. 4 is the basic set up for Rode's iteration. The iteration continues until $\delta$ falls below a predefined tolerance where we define $\delta = \max(|g_{i+1} - g_i|)$. We set the tolerance value as 0.1% of $\max(g_0)$, $g_0$ is defined from RTA. In Fig. B.1 we show different $g_i$s and the inset shows the rate of convergence which is exponential in nature.

## APPENDIX C: ANALYTICAL EXPRESSION FOR SEEBECK COEFFICIENT OF NON-DEGENERATE 2D MATERIALS

From Eq. (1) the current density $J_x$ can be written as

$$J_x = -\frac{e}{\Omega}\sum_k v g(k)\cos\theta \quad (C1)$$

This is because the symmetric part of the electronic distribution function does not contribute to any current. From Eq. (4) in absence of the in-scattering term $g(k)$ reduces to

$$g(k) = \frac{-\frac{eF}{\hbar}\frac{\partial f_0(k)}{\partial k} - v\frac{\partial f_0}{\partial x}}{S_{out}(k)+1/\tau_m(k)} \quad (C2)$$

Short circuiting the ends of the devices gives $F = 0$ and hence

$$g(k) = \frac{-v\frac{\partial f_0}{\partial x}}{\Gamma(k)} \quad (C3)$$

where $\Gamma(k) = S_{out}(k) + \frac{1}{\tau_m(k)}$ is the total out-scattering rate. Eq. (C3) also follows from RTA. Plugging this expression of $g(k)$ in Eq. (C1) and carrying out the 2D angular integration in $\boldsymbol{k}$−space we get

$$J_x = -\frac{4e\hbar}{\Omega E_N^{1-r}} \int E^{1-r} \frac{\partial f_0}{\partial x} dE \tag{C4}$$

We took $\Gamma \propto E^r$ and $E_N$ is just a normalization energy brought to preserve dimensions and does not affect the final results. We took care of spin and valley degeneracies though they are not important in current context. The job is to compute $\frac{\partial f_0}{\partial x}$. Using spatial charge homogeneity [19] ($\frac{\partial n}{\partial x} = 0$), and some algebraic manipulation we obtain for a non-degenerate 2D material

$$\frac{\partial E_F}{\partial x} = \frac{1}{T}\left(E_F - \frac{\int_0^\infty E e^{-\eta} dE}{\int_0^\infty e^{-\eta} dE}\right)\frac{\partial T}{\partial x} \tag{C5}$$

where $\eta = \frac{E-E_F}{k_B T}$, $E_F$ is the chemical potential, and $\frac{\partial T}{\partial x}$ is the applied temperature gradient. Eq. (C5), with a few more lines of algebra, reduces to $\frac{\partial E_F}{\partial x} = -k_B \frac{\partial T}{\partial x}$. Now for a non-degenerate semiconductor,

$$\frac{\partial f_0}{\partial x} = e^{-\eta} \frac{\partial \eta}{\partial x} \tag{C6}$$

and

$$\frac{\partial \eta}{\partial x} = -\frac{\eta}{T}\frac{\partial T}{\partial x} - \left(\frac{1}{k_B T}\right)\frac{\partial E_F}{\partial x} \tag{C7}$$

Plugging $\frac{\partial E_F}{\partial x}$ from Eq. (C5) in Eq. (C7) and using that in Eq. (C6) we get

$$\frac{\partial f_0}{\partial x} = \frac{e^{-\eta}}{T}(1-\eta)\frac{\partial T}{\partial x} \tag{C8}$$

Hence the current density becomes

$$J_x = -\frac{4e\hbar}{\Omega E_N^{1-r}} \frac{\partial T}{T \partial x} \int E^{1-r} e^{-\eta}(1-\eta) dE \tag{C9}$$

Now we look at Eq. 5(a) and see that $\sigma_x$ is the coefficient of $\frac{\partial E_F}{\partial x}$. Hence combining this view with Eq. (C4), (C6), and (C7) we get

$$\sigma_x = -\frac{4e^2 \hbar}{\Omega E_N^{1-r}} \frac{\partial T}{k_B T \partial x} \int E^{1-r} e^{-\eta} dE \tag{C10}$$

Using Eq. 5(a) and the definition of Seebeck coefficient, $\alpha = \frac{B_x}{\sigma_x T^2}$, where $B_x$ is a transport coefficient in Eq. 5(a), we can rewrite $\alpha$ as [19]

$$\alpha = -\frac{\left(\frac{J_x}{\sigma_x} - \frac{\partial \left(\frac{E_F}{e}\right)}{\partial x}\right)}{\frac{\partial T}{\partial x}} \tag{C11}$$

We use the previously obtained relation $\frac{\partial E_F}{\partial x} = -k_B \frac{\partial T}{\partial x}$ and Eq. (C9) and (C10) to calculate the right hand side of Eq. (C11). Usage of Gamma function leads to

$$\alpha = -\frac{k_B}{e}\left(2 - r - \frac{E_F}{k_B T}\right) \tag{C12}$$

Eq. (C12) is the 2D analogue of the expression given in [19] for bulk materials. An expression for $\alpha$ in a quantum well structure was given in [40], but it does not explicitly show the dependence of $\alpha$ on the scattering parameter $r$ as it most likely presumes $r = 0$ for a 2D system since the density of states is independent of energy. However, such assumption is not very accurate as our computation shows. A comparison of analytically calculated Seebeck coefficient for MoS$_2$ on SiO$_2$ at n$_s$ = $10^{11}$/cm$^2$ with that obtained with detailed numeric calculation is shown in Fig. C.1. The deviation in higher temperature is obvious due to the futility of RTA in presence of inelastic scattering. However, the analytical expression is useful in gaining intuition on $\alpha$ in a 2D material.

**FIGURES**

Fig. 1: Outline of the procedure to calculate the mobility and thermoelectric coefficients by solving the BTE and coupled current equations using Rode's method

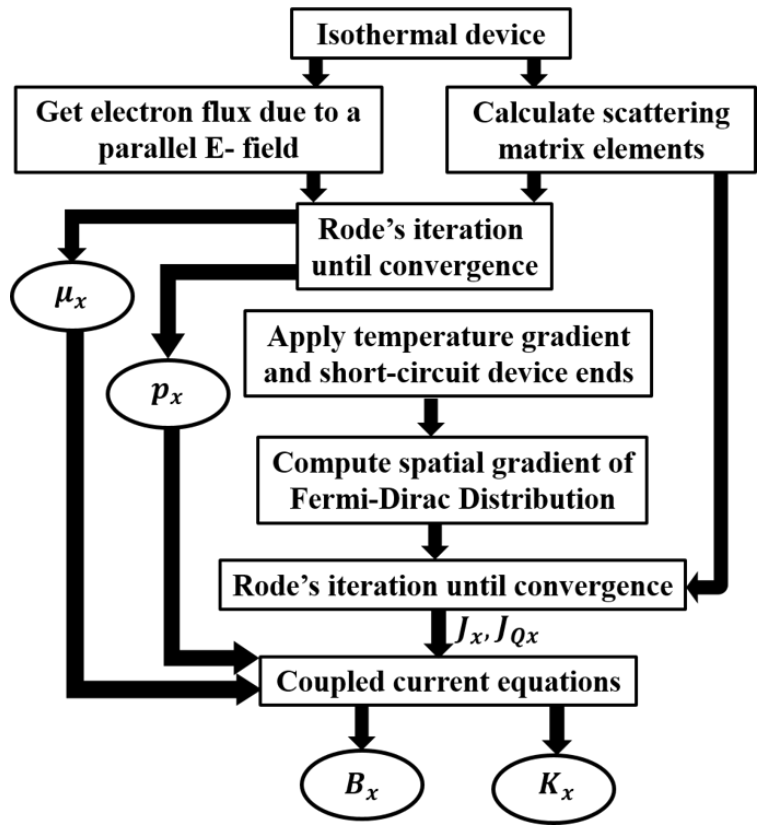

Fig. 2: (Color online) (a) Dispersion of the coupled IPP modes for $MoS_2$ on $HfO_2$ dielectric, (inset) device geometry and axes. (b) Phonon content in each IPP mode. The solid lines show content of first TO mode while the dashed ones show content of second TO mode. (c) Normalized scattering strength of the IPP modes due to the phonon content. The solid lines show scattering strength contribution of first TO mode while the dashed ones show the same for second TO mode. (d) The out-scattering rates ($S_{out}$) for electron-IPP interaction for two different electron densities for MoS2 on $HfO_2$. The corresponding impurity scattering rates are also shown for comparison.

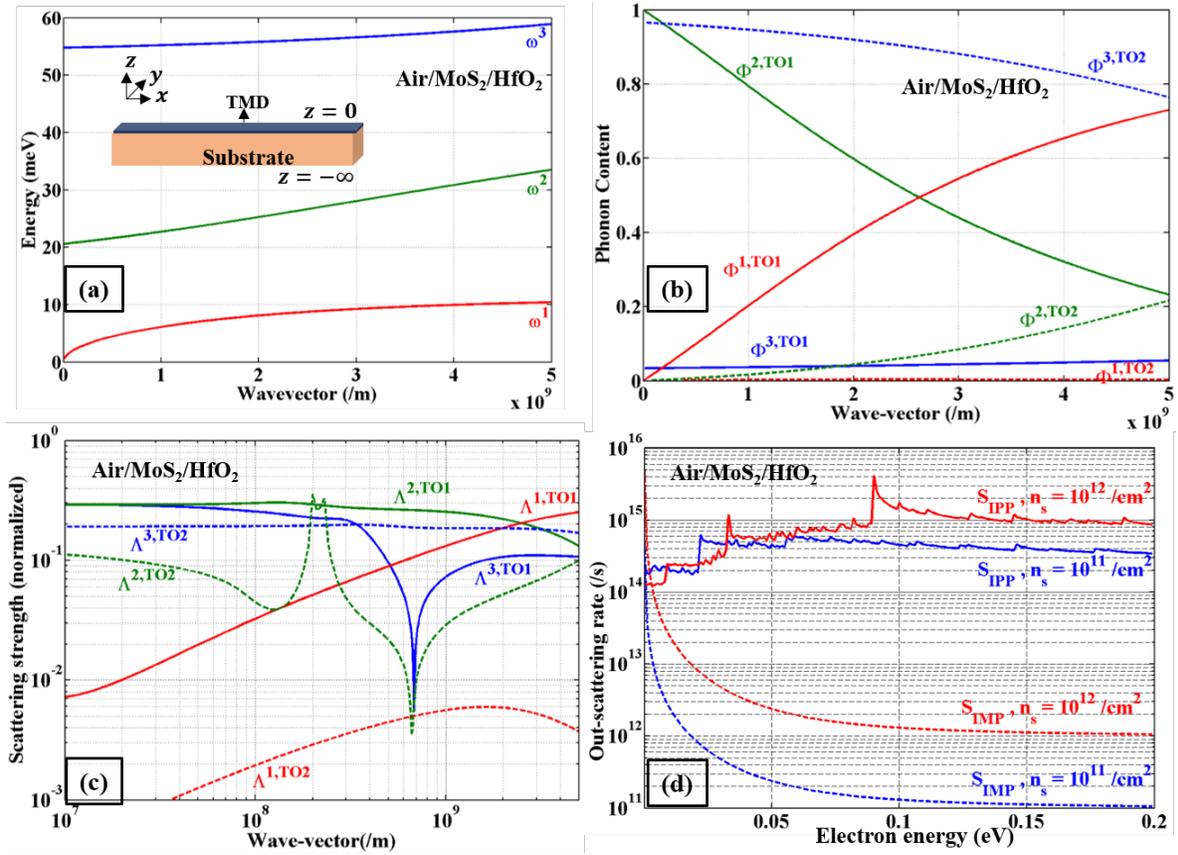

Fig. 3: (Color online) (a) Comparison of dynamic dielectric constant with the static Thomas-Fermi limit (inset) shows the contour plot of the dynamic dielectric function in the $(q, \omega)$ plane. (b) The single particle-excitation (SPE) boundaries – the red ones show boundaries for SP emission while the green one shows that for absorption. In the SPE region plasmon contribution to screening is reduced due to Landau damping. The undamped plasmon response is shown in black dashed line.

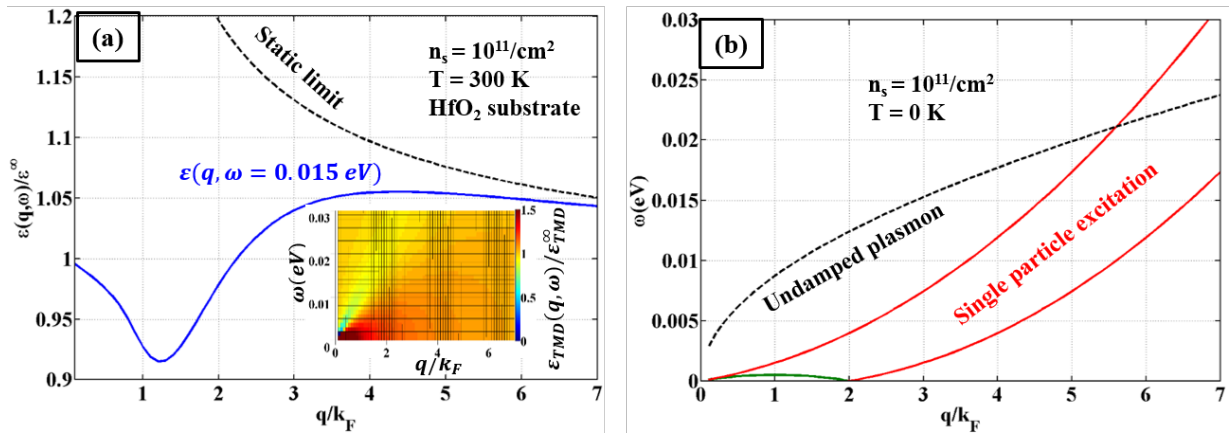

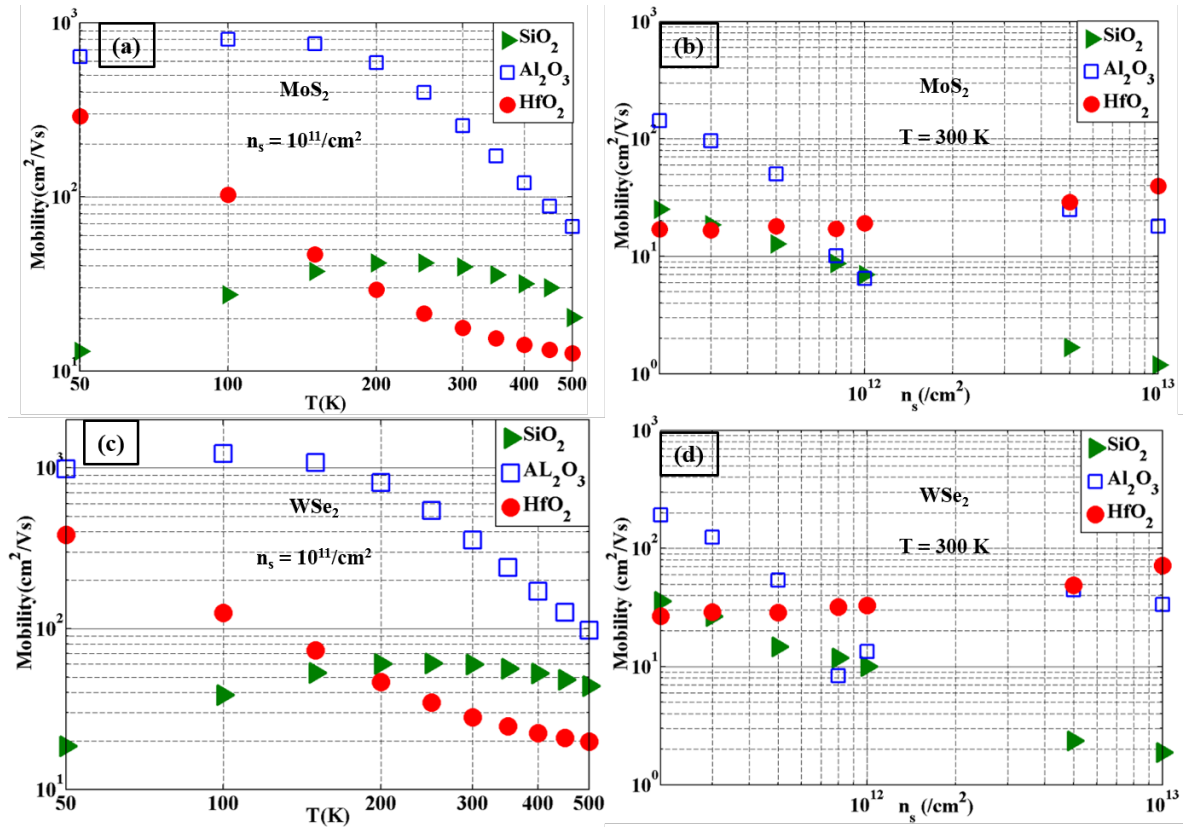

Fig. 4: (Color online) (a-b) Dependence of mobility for $MoS_2$ on temperature and electron density for different dielectric substrates; (c-d) similar plots for $WSe_2$

Fig. 5: (Color online) (a-b) Dependence of Seebeck coefficient for $MoS_2$ on temperature and electron density; (c-d) similar plots for $WSe_2$.

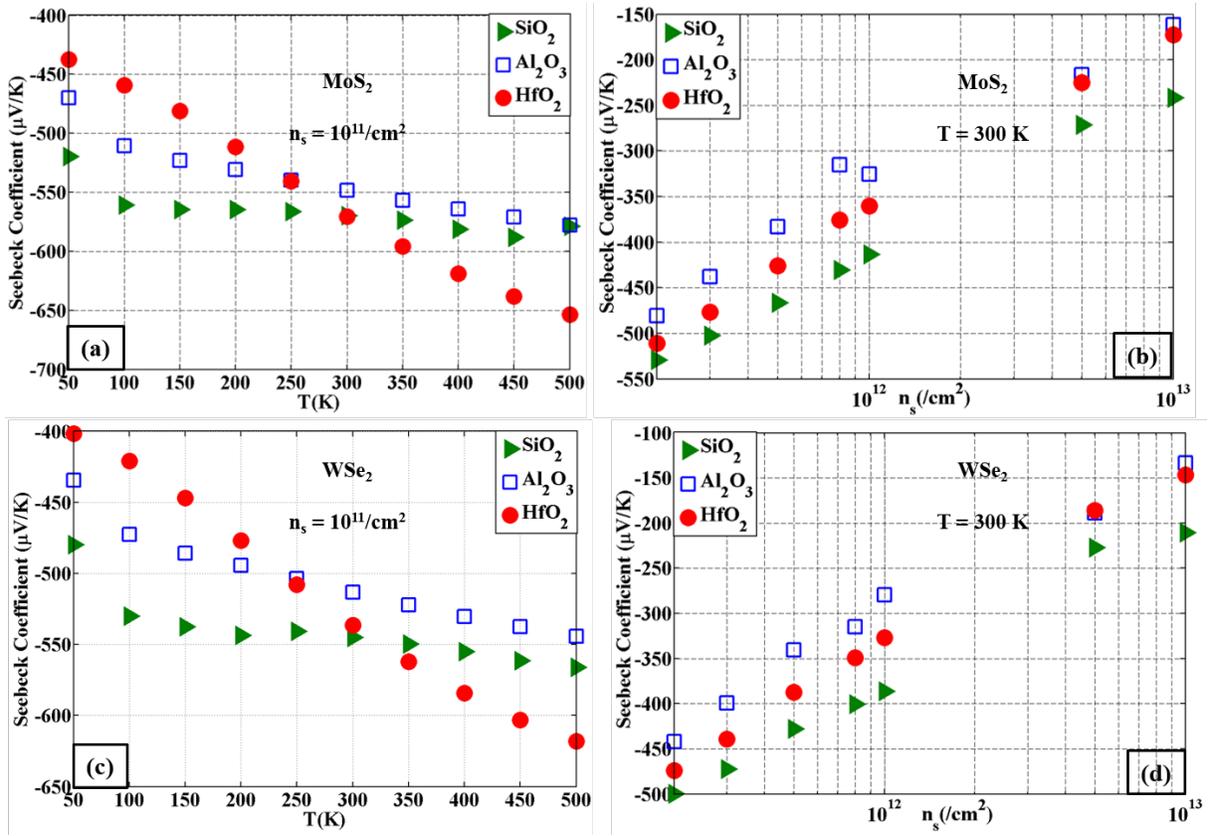

Fig. 6: (Color online) (a-b) Dependence of Peltier coefficient for $MoS_2$ on temperature and electron density; (c-d) similar plots for $WSe_2$.

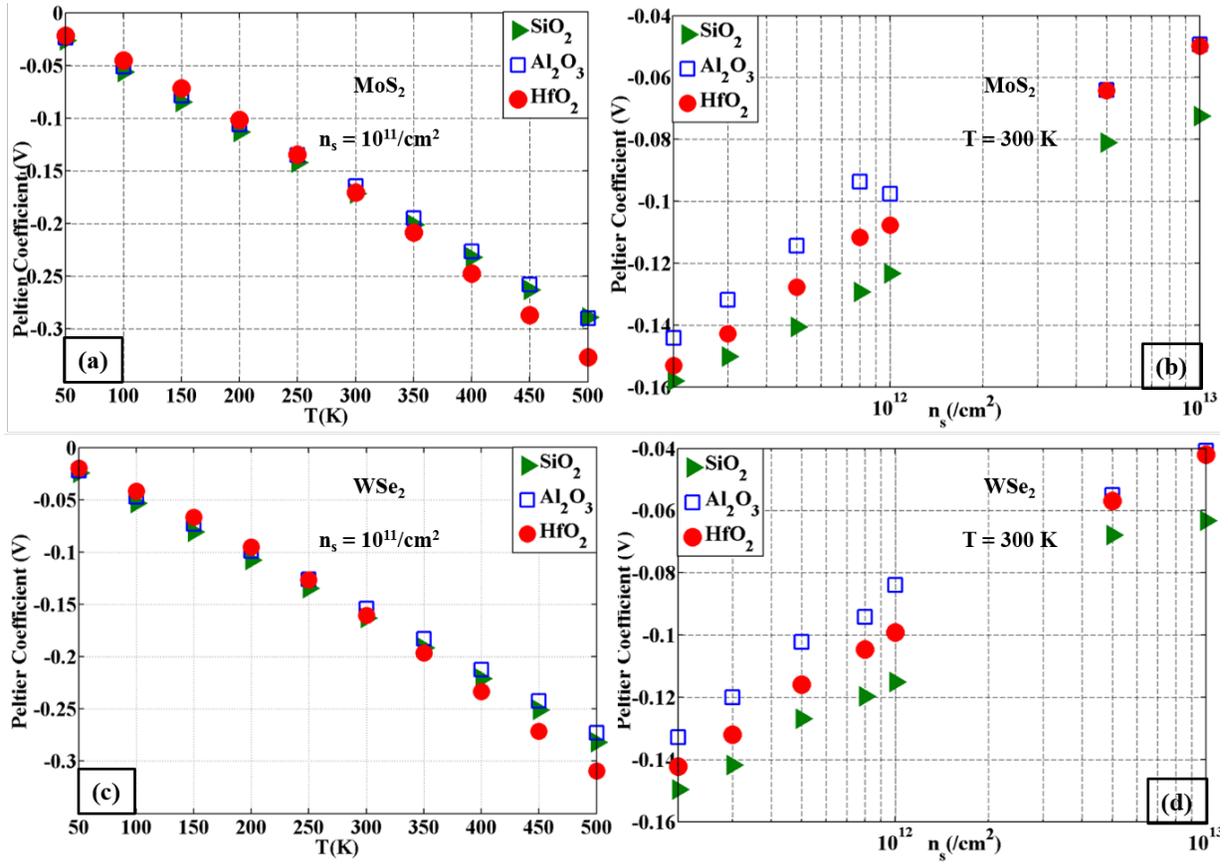

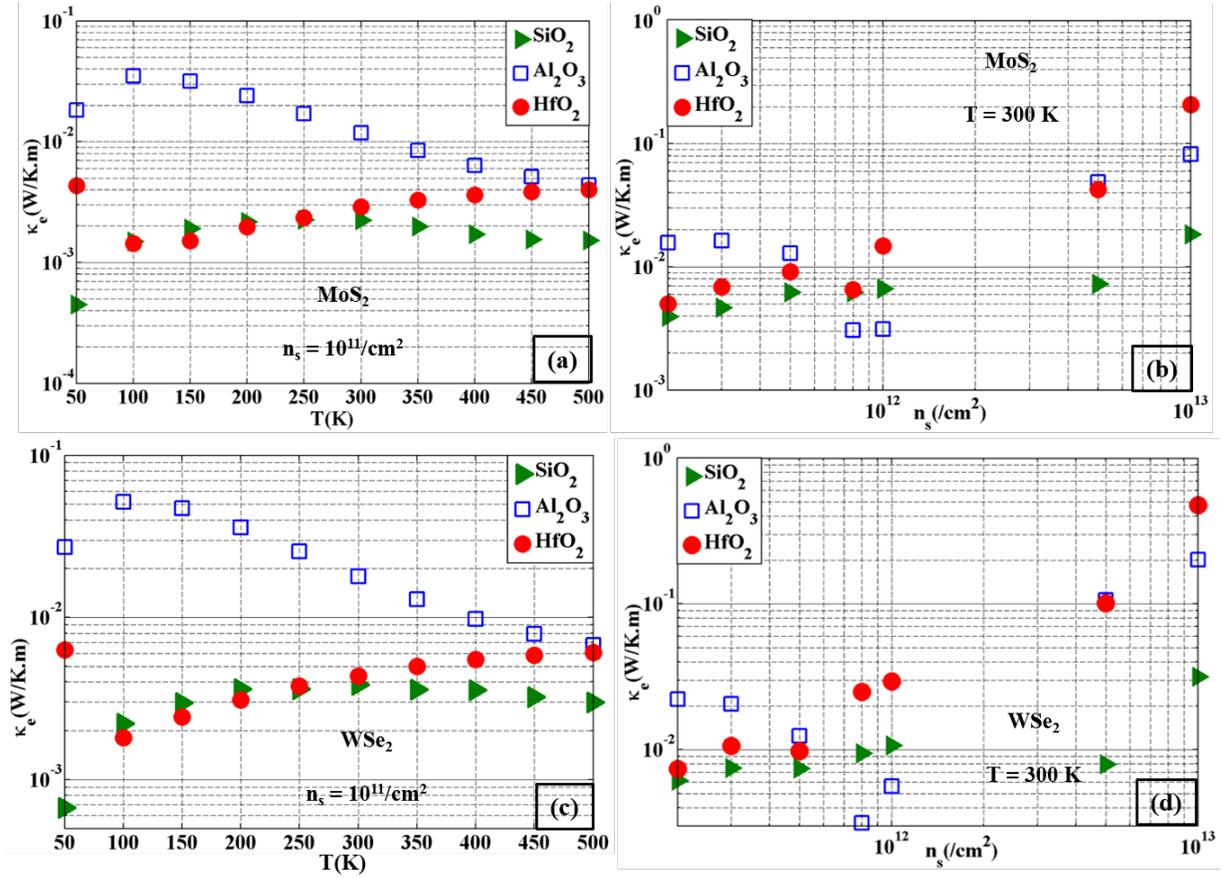

Fig. 7: (Color online) (a-b) Dependence of thermal conductivity (electronic) for $MoS_2$ on temperature and electron density for different dielectric substrates; (c-d) similar plots for $WSe_2$

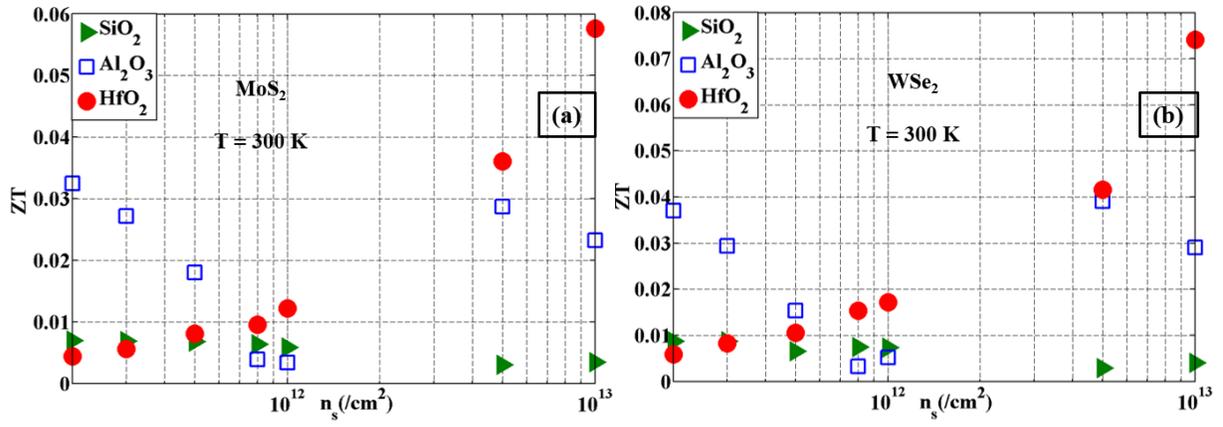

Fig. 8 (a) (Color online) Dependence of ZT for MoS$_2$ on electron density; (b) similar plots for WSe$_2$

Fig. B.1: Convergence of electronic distribution in Rode's iteration; (inset) rate of convergence with increasing iterations

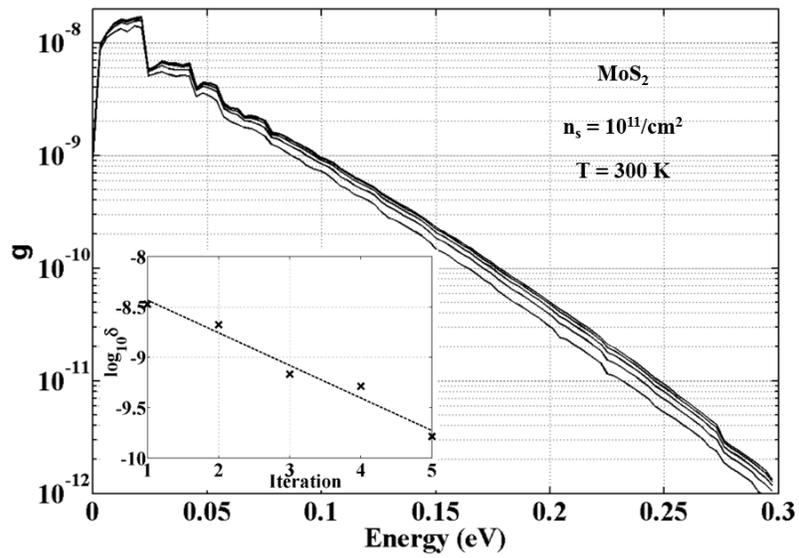

Fig. C.1: Comparison of analytically calculated Seebeck Coefficient with that calculated through detailed computation.

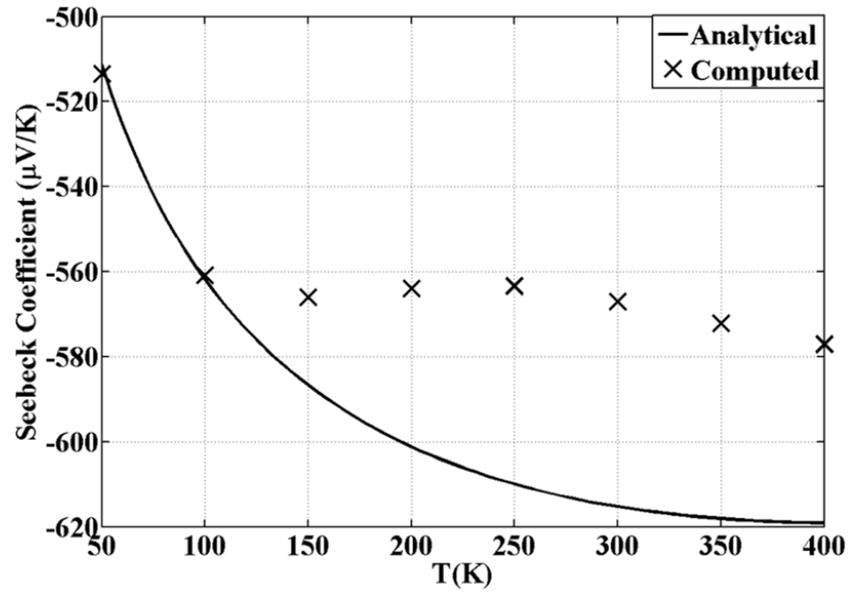

**TABLES**

Table I: TO-phonon and dielectric constants of different substrates (from [30])

|  | Substrate | | |
|---|---|---|---|
| Properties | SiO$_2$ | Al$_2$O$_3$ | HfO$_2$ |
| $\omega_{TO,1}$ (meV) | 55.6 | 48.18 | 12.40 |
| $\omega_{TO,2}$ (meV) | 138.1 | 71.41 | 48.35 |
| $\varepsilon_{sub}^{0}$ | 3.9 | 12.53 | 22 |
| $\varepsilon_{sub}^{int}$ | 2.8 | 6.58 | 7.27 |
| $\varepsilon_{sub}^{\infty}$ | 2.5 | 3.2 | 5.03 |

Table II: Properties of different TMDs ( [a]Ref [26] ; [b]Ref [41]; [c]Ref [29] ). [d]The high frequency dielectric constants of TMDs are taken to be equal to their static values due to negligible LO-TO splitting. Also these values for the two TMDs are taken to be equal. They are indeed very close to each other [42]

|  | TMD | |
|---|---|---|
| Properties | MoS$_2$ | WSe$_2$ |
| $\omega_{LO}$ (meV) | 48[a] | 32[b] |
| $m_e$ | 0.48[a] | 0.33[c] |
| $\varepsilon_{TMD}^{\infty}$ [d] | 4 | 4 |